\documentclass[12pt]{article}
\usepackage{fullpage}
\usepackage{amsmath}
\usepackage{amssymb}
\usepackage[dvips]{epsfig}

\def\##1{\underline{#1}}
\def\=#1{\underline{\underline{#1}}}

\def\+#1{\underline{\bf #1}}
\def\*#1{\underline{\underline{\bf #1}}}

\def\r#1{(\ref{#1})}
\def\l#1{\label{#1}}
\def\c#1{\cite{#1}}

\def\le{\left(}
\def\ri{\right)}
\def\les{\left[}
\def\ris{\right]}
\def\lec{\left\{}
\def\ric{\right\}}

\def\.{\mbox{ \tiny{$^\bullet$} }}

\def\epso{\epsilon_{\scriptscriptstyle 0}}

\def\muo{\mu_{\scriptscriptstyle 0}}

\def\ko{k_{\scriptscriptstyle 0}}

\def\Eo{\#E_{\, \scriptscriptstyle 0}}
\def\Ho{\#H_{\, \scriptscriptstyle 0}}

\def\ux{\hat{\#u}_x}
\def\uy{\hat{\#u}_y}
\def\uz{\hat{\#u}_z}

\begin{document}

\begin{center}

{\bf {\Large Electromagnetic waves with negative phase velocity
in Schwarzschild--de Sitter  spacetime }}

 \vspace{10mm} \large

Tom G. Mackay\footnote{Corresponding Author. Fax: + 44 131 650
6553; e--mail: T.Mackay@ed.ac.uk.} \\
{\em School of Mathematics,
University of Edinburgh, Edinburgh EH9 3JZ, UK}\\
\bigskip
 Akhlesh  Lakhtakia\footnote{Corresponding Author. Fax: +1 814 863
4319; e--mail: akhlesh@psu.edu; also
 affiliated with Department of Physics, Imperial College, London SW7 2 BZ,
UK}\\
 {\em Department of Engineering Science and
Mechanics\\ Pennsylvania State University, University Park, PA
16802--6812, USA}\\
\bigskip
Sandi Setiawan\footnote{Fax: + 44 131
650 6553; e--mail: S.Setiawan@ed.ac.uk.}\\
{\em School of Mathematics,
University of Edinburgh, Edinburgh EH9 3JZ, UK}\\

\end{center}

\begin{abstract}
The propagation of electromagnetic plane waves with negative phase
velocity (NPV) is considered in Schwarzschild--(anti--)de Sitter
spacetime. It is demonstrated that NPV propagation  occurs in
Schwarzschild--de Sitter spacetime at lower values of the
cosmological constant than is the case for de Sitter spacetime.
Furthermore, we report that  neither is NPV propagation observed in
Schwarzschild--anti--de Sitter spacetime, nor is  it
 possible outside the event horizon of a Schwarzschild blackhole.
\end{abstract}

\section{Introduction}
This communication concerns the propagation of electromagnetic
plane\-waves with \emph{negative phase velocity} (NPV) in curved
spacetime. The phase velocity is called negative if  the
time--averaged Poynting vector and the wavevector are oppositely
aligned \c{LMW}. The phenomenon of negative refraction~---~ which
has lately generated  considerable excitement in the
 electromagnetics and materials research communities \c{SSS}~---~follows
as a direct consequence of NPV propagation. The prospect of
technological applications, such as
 lenses with extremely low distortion, has prompted
intense efforts  by experimentalists and theoreticians directed
towards the development of NPV--supporting artificial
\emph{metamaterials} \c{pen2}.

The importance of  NPV propagation in astrophysical scenarios has
recently been emphasized. We  have previously shown that vacuum,
in association with certain spacetime metrics, can support NPV
propagation \c{LM04,MLS05}. For example, the metric of de Sitter
spacetime admits NPV propagation, whereas the anti--de Sitter
metric does not \c{MSL04}. Thus, an experimental means of
distinguishing between anti--de Sitter and de Sitter spacetimes is
offered via NPV propagation. Furthermore,
 regions  supporting NPV propagation are to be found
within the ergosphere of a rotating black hole \c{LMS05}. In this
context,  comparisons between NPV propagation and superradiant
scattering are noteworthy. While both phenomenons involve negative
energy densities, there are significant differences in terms of
directional properties and frequency bounds \c{SML05}.

In the present study we consider planewave propagation in
Schwarzschild--(anti--)de Sitter spacetime. Our analysis is based
on a formulation originally proposed by Tamm \c{Tamm}, wherein the
formal analogy between  electromagnetic propagation in
gravitationally affected vacuum and propagation in a (fictitious)
instantaneously responding medium is exploited \c{SS} .

\section{Schwarzschild--(anti--)de Sitter spacetime}
The static Schwarzschild--(anti--)de Sitter metric, with the
signature $(+,-,-,-)$, is conventionally expressed in spherical
coordinates as \c{gh1977,bh1998,pod1999,car2002,car2001}
\begin{equation}
ds^2 = (1-f)dt^2 - \frac{1}{1-f}\,dr^2 - r^2
(d\theta^2+\sin^2\theta \, d\phi^2) \,, \l{spher}
\end{equation}
wherein
$f =(2GMr^{-1} +  \Lambda r^2/3)/c^2$,
 $M$ is the mass of the black hole, $c$ is the speed of light
in vacuum in the absence of a gravitational field, $G$ is the
gravitational constant, and $\Lambda$ is the cosmological
constant. If the cosmological constant $\Lambda$ is positive, the
spacetime is called Schwarzschild--de Sitter spacetime, whereas
Schwarzschild--anti--de Sitter spacetime is characterized by
$\Lambda < 0$  \c{exact}. Let us note that \r{spher} reduces to
\begin{itemize}
\item[(a)] the Schwarzschild metric when $\Lambda = 0$, and
\item[(b)] the
cosmological metric (de Sitter or anti--de Sitter, according to
whether $\Lambda$ is positive or negative, respectively) when $M =
0$.
\end{itemize}
In terms of Cartesian coordinates $x = r \sin \theta \cos \phi$, $y =  r \sin \theta \sin \phi$,
and $z = r \cos \theta $,
the metric \r{spher} is  represented by  $g_{\alpha \beta}$  as\footnote{Roman indexes take the values 1,
2 and 3; while Greek indexes take the values 0, 1, 2, and 3.}
\begin{eqnarray}
\les \, g_{\alpha \beta} \, \ris & = &  \left( \begin{array}{cccc}
1-f & 0 & 0 & 0\\
\\
0 & -  1 - \frac{\displaystyle  fx^2}{\displaystyle r^2 \le 1-f \ri }
 & -\frac{\displaystyle  fxy}{\displaystyle r^2 \le 1-f \ri } & -\frac{\displaystyle  fxz}{\displaystyle r^2 \le 1-f \ri } \\
\\
0 & -\frac{\displaystyle  fxy}{\displaystyle r^2 \le 1-f \ri }
& -  1 - \frac{\displaystyle  fy^2}{\displaystyle r^2 \le 1-f \ri } & -\frac{\displaystyle  fyz}{\displaystyle r^2 \le 1-f \ri } \\
\\
0 & -\frac{\displaystyle  fxz}{\displaystyle r^2 \le 1-f \ri } &
-\frac{\displaystyle  fyz}{\displaystyle r^2 \le 1-f \ri } & -
 1
- \frac{\displaystyle  fz^2}{\displaystyle r^2 \le 1-f \ri }
\end{array} \right), \l{gab}
\end{eqnarray}
and its inverse $g^{\alpha \beta}$ as
\begin{eqnarray}
\les \, g^{\alpha \beta} \, \ris & = &  \left( \begin{array}{cccc}
\displaystyle{\frac{1}{1-f}} & 0 & 0 & 0\\
\\
0 & -  1 + \frac{\displaystyle  fx^2}{\displaystyle r^2  }
 & \frac{\displaystyle  fxy}{\displaystyle r^2  } & \frac{\displaystyle  fxz}{\displaystyle r^2  } \\
\\
0 & \frac{\displaystyle  fxy}{\displaystyle r^2  }
& -  1 + \frac{\displaystyle  fy^2}{\displaystyle r^2  } & \frac{\displaystyle  fyz}{\displaystyle r^2  } \\
\\
0 & \frac{\displaystyle  fxz}{\displaystyle r^2  } &
\frac{\displaystyle  fyz}{\displaystyle r^2  } & -
 1
+ \frac{\displaystyle  fz^2}{\displaystyle r^2  }
\end{array} \right). \l{gab_i}
\end{eqnarray}

Following common practice \c{SS,Skrotskii,Plebanski,Mashhoon} ,
the electromagnetic response of vacuum in curved spacetime may be
described by the constitutive relations of an equivalent,
instantaneously responding, medium   as per
\begin{equation}
\left.
\begin{array}{l}
\#D = \epso \=\gamma \. \#E
\\
\#B= \muo \=\gamma \. \#H
\end{array}
\right\}, \l{CR}
\end{equation}
wherein SI units are implemented.
Here,  $\epso = 8.854\times 10^{-12}$~F~m~$^{-1}$,
$\muo = 4\pi\times 10^{-12}$~H~m$^{-1}$,
and $\=\gamma$ is  the 3$\times$3 dyadic equivalent of the metric $\les \, \gamma_{ab} \, \ris $ with components
\begin{equation}
\gamma_{ab}  = -  \frac{g^{ab}}{g_{00}} .
\end{equation}

\section{Piecewise uniform approximation}
A global description of Schwarzschild--(anti--)de Sitter
spacetime is provided by  the constitutive relations \r{CR}.
 Let us partition the global spacetime into adjoining neighbourhoods.
At an  arbitrary location $\le \tilde{x}, \tilde{y}, \tilde{z} \ri$,
we consider the neighbourhood $\cal R$
 which is taken to be sufficiently small that
 the nonuniform
metric $\gamma_{ab}$ may  be reasonably approximated  by the
uniform metric $\tilde{\gamma}_{ab}$ \c{LMS05}. By stitching
together solutions from adjoining neighbourhoods, we formulate
 the global
solution.
 This piecewise
uniform approximation technique is commonly employed in solving
differential equations with nonhomogeneous coefficients
\c{Hoffman}. Thus, we have the uniform 3$\times$3 dyadic
representation
\begin{equation}
\={\tilde\gamma}  \equiv \les \, \tilde{\gamma}_{ab} \, \ris =  \frac{1}{1-\tilde{f}}
\left( \begin{array}{ccc}
  1-\frac{\displaystyle \tilde{f} \tilde{x}^2}{\displaystyle \tilde{r}^2}
 & -\frac{\displaystyle \tilde{f} \tilde{x}\tilde{y}}{\displaystyle \tilde{r}^2 } & -\frac{\displaystyle
\tilde{f} \tilde{x}\tilde{z}}{\displaystyle \tilde{r}^2 } \\
\\
-\frac{\displaystyle \tilde{f} \tilde{x}\tilde{y}}{\displaystyle \tilde{r}^2  } &  1-\frac{\displaystyle \tilde{f} \tilde{y}^2}{\displaystyle \tilde{r}^2}
 & -\frac{\displaystyle \tilde{f} \tilde{y}\tilde{z}}{\displaystyle \tilde{r}^2  }  \\
\\
-\frac{\displaystyle \tilde{f} \tilde{x}\tilde{z}}{\displaystyle \tilde{r}^2  }
& -\frac{\displaystyle \tilde{f} \tilde{y}\tilde{z}}{\displaystyle \tilde{r}^2 } &  1-\frac{\displaystyle \tilde{f} \tilde{z}^2 }{\displaystyle \tilde{r}^2}
\end{array} \right) \, \l{gamma_m}
\end{equation}
at  $\cal R$, with
\begin{equation}
\tilde{f} = \frac{2GM}{c^2 \tilde{r}} + \frac{\displaystyle
\Lambda \tilde{r}^2 }{\displaystyle 3c^2}
\end{equation}
and $
 \tilde{r}^2 =
\tilde{x}^2 + \tilde{y}^2 + \tilde{z}^2 $. We note that $
\mbox{det} \les \, \={\tilde\gamma} \, \ris = \le 1 - \tilde{f}
\ri^{-2}$.

\section{Plane waves in $\cal R$}
We seek planewave solutions
\begin{eqnarray}
&& \#E = {\rm Re} \lec\Eo \exp \les i \le  \#k \. \#r - \omega t
\ri \ris\ric\,, \l{pw_e} \qquad  \#H = {\rm Re} \lec\Ho \exp \les
i \le  \#k \. \#r - \omega t \ri \ris\ric\,,
\end{eqnarray}
 to the source--free Maxwell curl postulates
\begin{eqnarray}
&& \nabla \times \#E + \frac{\partial}{\partial t} \#B = \#0\,,
\qquad  \nabla \times \#H - \frac{\partial}{\partial t} \#D =
\#0\, \l{MP_2}
\end{eqnarray}
in $\cal R$. Here, the wavevector is denoted by $\#k$ and the
position vector  within the neighbourhood containing
$\le \tilde{x}, \tilde{y}, \tilde{z} \ri$ is represented by $\#r$;
 $\omega$ and $t$ are the angular frequency and time, respectively. Let us emphasize
 here that $\#r$ and $t$
 are   independent of $\cal R$.

The amplitudes $\Eo$ and $\Ho$ are complex--valued with
  $i=\sqrt{-1}$.
Upon combining \r{pw_e} and \r{MP_2}, we find after  some
algebraic manipulation that
\begin{equation}
\=W \. \Eo = \#0 \,, \l{ev_eq}
\end{equation}
where
\begin{equation}
\=W = \le \ko^2 \mbox{det} \les \, \={\tilde\gamma} \, \ris -
\#k\. \={\tilde\gamma} \. \#k \ri \, \=I + \#k\,\#k \.
\={\tilde\gamma}\,,
\end{equation}
and  $\ko = \omega \sqrt{\epso \muo}$. Thus, the dispersion
relation ${\rm det} \les \, \=W \, \ris = 0$ emerges, which can be
recast as
\begin{equation}
\ko^2\,  \mbox{det} \les \, \={\tilde\gamma} \, \ris
 \le \ko^2 \,\mbox{det} \les \, \={\tilde\gamma} \, \ris - \#k\. \={\tilde\gamma} \. \#k \ri^2 = 0\,.
\end{equation}
Clearly,
 the wavevectors
must satisfy the condition
\begin{equation}
\#k\.\={\tilde\gamma}\.\#k = \ko^2 \, \mbox{det} \, \les \, \={\tilde\gamma} \,
\ris\,, \l{disp_cond}
\end{equation}
if  $\={\tilde\gamma}$ is nonsingular.

Let us consider now the eigensolutions of \r{ev_eq}. In view  of
\r{disp_cond}, we have
\begin{equation}
\#k \, \#k \.  \={\tilde\gamma} \. \Eo = \#0 \,; \l{ev_cond}
\end{equation}
hence, it follows that $\Eo$ is  orthogonal to $\#k \.
\={\tilde\gamma}$. The spacetime metric described by \r{spher} is
spherically symmetric. Therefore, there is no loss of generality
in choosing the wavevector
\begin{equation}
\#k = k \uz\, ,
\end{equation}
with
 $\uz$
being the
unit vector
lying along the $z$  Cartesian axis.
Thereby,
\begin{equation} \#k \. \={\tilde\gamma} =
k(\tilde{\gamma}_1 \ux + \tilde{\gamma}_2 \uy +
\tilde{\gamma}_3\uz)\,,
\end{equation}
where
\begin{equation}
\tilde{\gamma}_1 = - \frac{\displaystyle \tilde{f}
\tilde{x}\tilde{z}}{ \displaystyle \le 1-\tilde{f} \ri
\tilde{r}^2} \, ,\quad \tilde{\gamma}_2 = - \frac{\displaystyle
\tilde{f} \tilde{y}\tilde{z}}{\displaystyle \le 1-\tilde{f} \ri
\tilde{r}^2} \, ,\quad \tilde{\gamma}_3 = \frac{\displaystyle
\tilde{f}_z}{\displaystyle \le 1-\tilde{f} \ri}\,,
\end{equation}
with
\begin{equation}  \tilde{f}_z = 1-\frac{ \tilde{f}
\tilde{z}^2 }{ \tilde{r}^2}\,,
\end{equation}
and $\ux$ and $\uy$ being unit vectors  lying along the $x$ and $y$  Cartesian
axes, respectively.

The two  linearly independent eigenvectors
\begin{eqnarray}
&& \#e_{\,1} = \tilde{\gamma}_2 \ux - \tilde{\gamma}_1 \uy \,,
\qquad  \#e_{\,2} = \tilde{\gamma}_1 \tilde{\gamma}_3 \ux
+\tilde{\gamma}_2 \tilde{\gamma}_3 \uy - \le \tilde{\gamma}_1^2 +
\tilde{\gamma}_2^2 \ri \uz \, \l{e2}
\end{eqnarray}
satisfy \r{ev_cond};  hence, we have the general solution
\begin{equation}
\Eo = C_1 \#e_{\,1} + C_2 \#e_{\,2}
\,, \l{E_vec}
\end{equation}
with $C_1$ and $C_2$ being arbitrary complex--valued constants.
The corresponding  expression for $\Ho$ follows  from the Maxwell
postulates as
\begin{equation}
\Ho = \frac{k}{\omega \muo}
 \, \les \, C_1 \,
\le 1-\tilde{f} \ri \#e_{\,2} - C_2 \tilde{f}_z \#e_{\,1}  \, \ris
 \l{H_vec} \,.
\end{equation}

To calculate the  wavenumbers we turn to the  dispersion equation
\r{disp_cond}. For $\#k$ aligned with $\hat{\#u}_{z}$  we obtain
the $k$--quadratic expression
\begin{equation}
k^2 \frac{\displaystyle \tilde{f}_z}{\le 1-\tilde{f} \ri} -
\frac{\displaystyle \ko^2}{\displaystyle \le 1-\tilde{f} \ri^2} = 0 \,,
\l{quad}
\end{equation}
from which  the wavenumbers
\begin{eqnarray}
k &=&  \pm \ko \, \les \le 1-\tilde{f} \ri \tilde{f}_z  \ris^{-1/2}
 \end{eqnarray}
straightforwardly emerge.

The requirement that $k \in \mathbb{R}$ imposes the condition
\begin{equation}
\le 1-\tilde{f} \ri \tilde{f}_z > 0\,; \l{mm_cond}
\end{equation}
in other words,
 both $\le 1-\tilde{f} \ri$ and $\tilde{f}_z$ must have the same signs for
 propagating planewave solutions.

\section{NPV Condition}

The propagation of planewaves with NPV is signalled by the
inequality \c{MLS05,LMS05}
\begin{equation}
\#k \. \langle \, \#P \, \rangle_t < 0\,,
\end{equation}
where
$\langle \, \#P \, \rangle_t =(1/2)\, {\rm Re}\lec
\Eo\times \Ho^\ast\ric
$
is the the time--averaged Poynting vector. The general solution
\r{E_vec} and \r{H_vec} delivers
\begin{equation}
\langle \, \#P \, \rangle_t = \frac{k}{2\,\omega \muo} \les \, |
C_1 |^2 \le 1-\tilde{f} \ri + | C_2 |^2 \tilde{f}_z \, \ris \,
\#e_{\,1} \times \#e_{\,2}\,.
\end{equation}
By virtue of \r{e2},
the orientation of  $\langle \, \#P \, \rangle_t$ is provided by
the vector
\begin{equation}
\#e_1 \times \#e_2 = \le  \tilde{\gamma}^2_1 + \tilde{\gamma}^2_2
\ri  \le \tilde{\gamma_1} \ux + \tilde{\gamma_2} \uy +
\tilde{\gamma_3} \uz  \ri.
\end{equation}
Thus,
\begin{equation}
\#k \. \le  \#e_1 \times \#e_2 \ri = k \frac{\tilde{f}_z}{
1-\tilde{f} }  \les \, \frac{ \tilde{f} \tilde{z}}{ \tilde{r}^2
\le 1-\tilde{f} \ri} \, \ris^2 \le \tilde{x}^2 + \tilde{y}^2
\ri\,,
\end{equation}
and it follows that
\begin{eqnarray}
\#k \. \langle \, \#P \, \rangle_t &=& \frac{1}{2\, \omega \muo}
\les \, \frac{k \tilde{f} \tilde{z}}{\tilde{r}^2 \le 1-\tilde{f}
\ri} \, \ris^2 \le \tilde{x}^2 + \tilde{y}^2 \ri \le \, | C_1 |^2
+ | C_2 |^2 \frac{\tilde{f}_z}{ 1-\tilde{f} } \, \ri
\tilde{f}_z\,.
\end{eqnarray}

Hence, by exploiting \r{mm_cond}, we see that  NPV arises as a
consequence of  $\tilde{f}_z < 0$ ; i.e., NPV propagation occurs
provided that
\begin{equation}
\frac{6 G M }{\tilde{r}^3} + \Lambda
> \frac{3 c^2}{\tilde{z}^2}\, . \l{NPV}
\end{equation}
Let us emphasize that while \r{NPV} has been derived for the
neighbourhood $\cal R$, the location of $\cal R$ is arbitrary
within Schwarzschild--(anti--)de Sitter spacetime. Therefore, the
NPV inequality \r{NPV}  applies generally.

In order to contextualize  the NPV condition \r{NPV}, it is helpful to introduce
the event horizon for a Schwarzschild black hole
 which lies at $r = r_{\mbox{\tiny{Sch}}} = 2 GM/ c^2$
and the event horizon for de Sitter spacetime
 which lies at $r = r_{\mbox{\tiny{deS}}} = c \sqrt{3/\Lambda}$ \c{Inverno}.
Thereby, we see that \r{NPV} yields the following {\em sufficient\/}
conditions for NPV propagation:
\begin{equation}
\left. \begin{array}{l} \tilde{r}^3 > r^2_{\mbox{\tiny{deS}}} \le
\tilde{r} - r_{\mbox{\tiny{Sch}}} \ri \qquad \mbox{for} \quad
\Lambda > 0 \\ \\\tilde{r}^3 < \displaystyle{\frac{3 c^2}{\Lambda}
\le \tilde{r} - r_{\mbox{\tiny{Sch}}} \ri \qquad \mbox{for} \quad
\Lambda < 0}
\end{array}
\right\}\,. \l{EH}
\end{equation}

\section{Concluding remarks}

The general condition \r{NPV} has been derived for NPV propagation
in Schwarzschild--(anti--)de Sitter spacetime. Analysis of this condition allows
us to make the following conclusions:
\begin{itemize}
\item[(a)] If $\Lambda =0$ (i.e., the spacetime is described by the Schwarzschild metric),
then NPV propagation is indicated by the inequality
\begin{equation}
\frac{r_{\mbox{\tiny{Sch}}} }{ \tilde{r}}
> \frac{\tilde{r}^2}{\tilde{z}^2}\, . \l{NPV_Sch}
\end{equation}
Therefore,  we deduce that NPV propagation is  not possible
 outside the event horizon of a Schwarzschild black hole (i.e., for $ \tilde{r}
> r_{\mbox{\tiny{Sch}}}$).
\item[(b)]  If $M=0$ (i.e., the spacetime is (anti--)de Sitter), then
NPV is signalled by the inequality
\begin{equation}
\Lambda
> \frac{3 c^2}{\tilde{z}^2}\,,  \l{NPV_deS}
\end{equation}
in agreement with earlier results \c{MSL04}; i.e.,  anti--de
Sitter spacetime does not support NPV propagation. Furthermore, we
observe that it is the region of spacetime which lies outside the
de Sitter event horizon which supports  NPV propagation.

Let us analyze the NPV inequality \r{NPV_deS} in view of the
uniform approximation implemented in considering the neighbourhood
$\cal R$. The linear dimensions $\delta$ of $\cal R$ are taken to
be small relative to the global spacetime curvature and large
compared to the electromagnetic wavelengths as given by $2 \pi/k$.
Since the Ricci scalar $R$~---~which provides a measure of the
inverse radius of spacetime curvature squared~---~is given by $R =
4 \Lambda / c^2$ for de Sitter spacetime \c{Peebles}, we have
\begin{equation}
\frac{2 \pi}{| k |}  \ll \delta \ll \frac{c}{2} \sqrt{\frac{\rho}
{| \Lambda | }} \,, \l{partition}
\end{equation}
with $\rho$ being a proportionality constant. Hence,  the
partition of global spacetime requires
\begin{equation}
| \Lambda | \ll  \frac{c^2 | k |^2 \rho }{16 \pi^2}\,. \l{p2}
\end{equation}
The inequalities \r{NPV_deS} and \r{partition} are mutually
compatible as the
 linear dimensions $\delta$ of the neighbourhood
${\cal R}$ are chosen independently of the  $\tilde{z}$ coordinate
specifying the location of  ${\cal R}$.

\item[(c)] By
comparing \r{NPV} and  \r{NPV_deS}, it is clear that in
Schwarzschild--de Sitter spacetime NPV propagation occurs for
smaller (positive) values of $\Lambda$ than is the case for de
Sitter spacetime.

\item[(d)] Our numerical investigations have shown that the NPV inequality
\r{NPV} is not satisfied for $\Lambda < 0$  (i.e., Schwarzschild--anti--de Sitter
spacetime) in physically probe--able regions of spacetime.
\end{itemize}

\vspace{8mm}

\noindent {\bf Acknowledgments:}
 SS acknowledges EPSRC for support under grant GR/S60631/01. As often, AL thanks the Mercedes Foundation
 for continuous support.

\end{document}